\documentclass[graybox]{svmult}

\usepackage{mathptmx}       
\usepackage{helvet}         
\usepackage{courier}        
\usepackage{type1cm}        
\usepackage{makeidx}         
\usepackage{graphicx}        
\usepackage{multicol}        
\usepackage[bottom]{footmisc}

\def\la{\mathrel{\mathpalette\fun <}}
\def\ga{\mathrel{\mathpalette\fun >}}
\def\fun#1#2{\lower3.6pt\vbox{\baselineskip0pt\lineskip.9pt
  \ialign{$\mathsurround=0pt#1\hfil##\hfil$\crcr#2\crcr\sim\crcr}}}

\makeindex             

\begin{document}

\title*{Interdisciplinary Aspects of High-Energy Astrophysics}
\author{G\"unter Sigl}
\institute{G\"unter Sigl \at II. Institut f\"ur theoretische Physik, Universit\"at Hamburg,
Luruper Chaussee 149, D-22761 Hamburg, Germany  \email{guenter.sigl@desy.de}}
%
%
\maketitle

\abstract*{Modern astrophysics, especially at GeV energy scales and above
is a typical example where several disciplines meet: The location and distribution
of the sources is the domain of astronomy. At distances corresponding to significant
redshift cosmological aspects such as the expansion history come into play. Finally,
the emission mechanisms and subsequent propagation of produced high energy particles
is at least partly the domain of particle physics, in particular if new phenomena
beyond the Standard Model are probed that require base lines and/or energies
unattained in the laboratory. In this contribution we focus on three examples:
Highest energy cosmic rays, tests of the Lorentz symmetry and the search for new
light photon-like states in the spectra of active galaxies.}

\abstract{Modern astrophysics, especially at GeV energy scales and above
is a typical example where several disciplines meet: The location and distribution
of the sources is the domain of astronomy. At distances corresponding to significant
redshift cosmological aspects such as the expansion history come into play. Finally,
the emission mechanisms and subsequent propagation of produced high energy particles
is at least partly the domain of particle physics, in particular if new phenomena
beyond the Standard Model are probed that require base lines and/or energies
unattained in the laboratory. In this contribution we focus on three examples:
Highest energy cosmic rays, tests of the Lorentz symmetry and the search for new
light photon-like states in the spectra of active galaxies.}

\section{Introduction}
\label{sec:1}
High energy astrophysics is nowadays a very interdisciplinary research
field which either uses input from or provides new output to other
fields including astronomy, cosmology, particle physics and even
philosophy and (astro)biology. Examples very this becomes especially
obvious includes the use of active galactic nuclei to probe the
formation of structure at very high redshift of order ten, high energy
cosmic rays as probes for the annihilation or decay of dark matter and
the use of ``standard candles'' such as exploding white dwarfs and
(more recently) gamma-ray bursts to probe the expansion history of the
Universe.

A particular problem that sometimes occurs at these intersections are
different languages spoken by the different communities. In general,
however, a lot of progress has been made in that respect. This is the
case in particular in astroparticle physics, a still young but meanwhile well
established research discipline in its own right. This can be seen not
least from the fact that funding agencies in most countries have
developed programs and instruments aiming in specifically at this field.

The present paper can naturally cover
at most a tiny fraction of interesting examples for such interfaces
between neighboring research fields.
We specifically focus on three topics at the interface
beween astronomy, high energy astrophysics and particle physics:
First, ultra-high energy cosmic rays, traditionally understood
as particles with energies above $10^{18}\,$eV, have been observed
with energies up to a few times $10^{20}\,$eV, which is a macroscopic
energy of about 50 Joules, presumably of just one elementary
particle. Therefore,
very likely, the sources of these ultra-energetic particles have to be
exceptionally powerful and visible in other wavelengths and
channels. The search of these sources has thus a strong relation to astronomy.

Second, the macroscopic energies of these particles makes them natural
test beams for particle physics at energies that cannot be achieved in
the laboratory in the foreseeable future. In particular, tiny
violations of fundamental symmetries of Nature, such as the Lorentz
symmetry, may become magnified at large energies. We are still lacking
a description of gravity that is consistent with quantum mechanics and
the way gravity unifies with the electromagnetic, weak and strong
interactions may only manifest itself at energies approaching the
Planck scale. In this case, high energy astrophysics may be an
indispensable tool for the phenomenology of quantum gravity.

Finally, at the opposite, low energy end, new physics may also exist
in the form of very light particles that may morph into photons and
vice versa. The strongest constraints on such possibilities that are
often motivated by models of fundamental physics such as string theory
and loop quantum gravity often come from astrophysical and
cosmological observations which offer the largest baselines and the
highest energies.

\section{Astronomy with the Highest Energy Particles of Nature ?}
\label{sec:2}
The research field of ultra-high energy cosmic rays started 
in 1938 when Pierre Auger proved the existence of
extensive air showers (EAS) caused by primary particles
with energies above $10^{15}\,$eV by simultaneously observing
the arrival of secondary particles in Geiger counters many meters
apart~\cite{auger_disc}. Since that time, ultra-high energy cosmic
rays (UHECRs)
have challenged the imagination of physicists and astrophysicists alike.
The first cosmic ray with energy above $10^{20}\,$eV was discovered by
John Lindsley in 1963 at the Volcano Ranch Observatory~\cite{Linsley:1963km}.
The record holder is probably still the famous ``Fly's Eye event'' of
$\simeq3\times10^{20}\,$eV~\cite{Bird:1994uy} and quickly, scientists were looking for
astronomical sources~\cite{Elbert:1994zv}.
Around the same time, the Akeno Giant Air Shower Array (AGASA) caused
excitement because
it observed an UHECR spectrum continuing seemingly as a power law around
$10^{20}\,$eV. This was contrary to expectations because the famous
Greisen-Zatsepin-Kuzmin (GZK) effect~\cite{gzk} predicts that nucleons loose
their energy within about 20 Mpc above a threshold of
$\simeq6\times10^{19}\,$eV~\cite{stecker}
due to pion production on the cosmic microwave background which is a
relic of the early Universe. As long as we do not live in a strong
over-density of UHECR sources, this would predict a strong suppression
of the UHECR flux above that threshold, often somewhat misleadingly called
the ``GZK cutoff''. Meanwhile, a flux suppression consistent with the
GZK effect has been observed by the more recent High Resolution Fly's
Eye~\cite{hires-spec}
and Pierre Auger~\cite{auger-spec} experiments and it is likely that the AGASA
spectrum was due to an overestimate of the UHECR energies. 

These more recent, higher statistics data, however, raised other, no less
interesting questions: For the first time, the Pierre Auger
Observatory which observes the Southern hemisphere from Argentina
has accumulated enough statistics at the highest energies to see signs
of anisotropy: A significant correlation with the 12th edition of the
V\'eron-Cetty and V\'eron catalog of nearby AGNs was observed for
events with energies above $56\,$EeV~\cite{auger-anisotropy}. This is
very suggestive because it is also the energy scale above which the
GZK effect limits the range of primary cosmic rays to $\sim50\,$Mpc.
This does not necessarily mean that these objects represent the sources,
but it suggests that the real UHECR sources follow an anisotropic distribution
that is similar to nearby AGNs. This may not be surprising if the sources
are astrophysical accelerators which follow the local large scale structure.
Unfortunately, with accumulation of more data, these correlations have
weakened~\cite{correlation:2010zzj}. The fraction of events above
55 EeV correlating with the Veron Cetty Catalog has came down from
$69^{+11}_{-13}$\% to $38^{+7}_{-6}$\% compared to 21\% expected
for isotropy. If one divides the sky distribution into a component
correlating, for example, with the 2MASS redshift survey
and an isotropic component, this corresponds to a relatively large
isotropic fraction of 60--70\%~\cite{correlation:2010zzj}. Still, an
excess of correlations is seen with 2MASS redshift survey
at 95\% confidence level. On the other hand, in the Northern
hemisphere, the HiRes
experiment has not seen any correlations~\cite{Abbasi:2008md}.

The nature and location of UHECR
sources is thus still an open question in which general theoretical
considerations play a significant role.
Accelerating particles of charge $eZ$ to an energy $E_{\rm max}$ requires
an induction ${\cal E} \ga E_{\rm max}/(eZ)$. With $Z_0\simeq100\,\Omega$
the vacuum impedance, this requires dissipation of a minimal power
of~\cite{Lovelace,Blandford:1999hi}
\begin{equation}\label{eq:Lmin}
  L_{\rm min}\simeq\frac{{\cal E}^2}{Z_0}\simeq 10^{45}\,Z^{-2}\,
  \left(\frac{E_{\rm max}}{10^{20}\,{\rm eV}}\right)^2\,
  {\rm erg}\,{\rm s}^{-1}\,.
\end{equation}
When expressing the square of the product of the magnetic field in an
accelerator with its size in terms of a luminosity, this condition can
be expressed in terms of the Hillas-criterium~\cite{hillas-araa} which
states that the gyro-radius of a charged particle at the maximal
acceleration energy must fit within the accelerator. Eq.~(\ref{eq:Lmin})
suggests that the power requirements are considerably relaxed for
heavier nuclei which is easy to understand because an estimate solely
based on motion of charged particles in magnetic fields can only
depend on their rigidity $E/Z$. However, the Hillas criterion and
Eq.~(\ref{eq:Lmin}) are necessary but in general not sufficient since
they do not take into
account energy loss processes within the source. Extensions of the
conditions on UHECR sources that include  energy-loss processes have
recently been discussed in Ref.~\cite{Ptitsyna:2008zs}. An interesting
argument linking UHECR sources to their luminosity at radio
frequencies has been put forward by
Hardcastle~\cite{Hardcastle:2010cq}. He concludes that if UHECRs are
predominantly protons, then very few sources should
contribute to the observed flux. These sources should be easy to
identify in the radio and their UHECR spectrum should cut off steeply
at the observed highest energies. In contrast, if the composition is
heavy at the highest energies then many radio galaxies could
contribute to the UHECR flux but due to the much stronger deflection
only the nearby radio galaxy Centaurus A may be identifiable.

In fact, the Pierre Auger data reveal a clustering of
super-GZK events towards the direction of Centaurus
A (NGC 5128)~\cite{Moskalenko:2008iz,correlation:2010zzj}, whereas
other directions on the sky with an overdensity of potential UHECR
accelerators such as the Virgo cluster containing the prominent radio
galaxy M87 show an apparent deficit in such
events~\cite{Gorbunov:2007ja}. This is somewhat surprising since,
although Cen A is the closest radio galaxy and the third-strongest
radio source in the sky, it is a relatively weak elliptical
radio galaxy~\cite{Rieger:2009pm}, making it difficult to reach the
required UHECR energies. However, one should note that the UHECR
events observed towards
Cen A could at least partly originate from sources within the
Centaurus galaxy cluster which is located just behind Cen A and is
itself part of the Hydra-Centaurus supercluster. In
any case, due to its closeness, Cen A has been observed in many
channels. For example, its  lobes have been detected in 200 MeV
gamma-rays by Fermi LAT~\cite{cenA-lobes-fermi}, and its core was
observed by Fermi LAT~\cite{cenA-core-fermi}. These observations and
its potential role as a major local UHECR accelerator has
lead to many multi-messenger model building efforts for Cen
A~\cite{Rieger:2009pm,cenA-tomas}.
As an example, in Ref.~\cite{cenA-tomas} it was pointed out that
proton acceleration
in the jet of Cen A is hard to reconcile with Cen A observations in
TeV gamma-rays by HESS ~\cite{cenA-hess} if gamma-rays are produced
by proton-proton interactions. Instead, p$-\gamma$ interactions in the core are
consistent with these observations.

We note in passing that another potential UHECR source are gamma-ray
bursts (GRBs)~\cite{Dermer:2010km}. Although GRBs individually have
more than adequate power to achieve the required maximal acceleration
energies, but may be disfavored in terms of local power density compared
to an UHECR origin in AGNs and radio galaxies.

Another interesting new question concerns the chemical composition
of highest energy cosmic rays: The depth in the atmosphere where
particle density in the giant air showers observed by the Pierre Auger
Observatory is maximal, and in particular the fluctuations of the
depth of shower maximum from event to event, when compared with air
shower simulations, point towards a heavy
composition for energies $10^{19}\,{\rm eV}\la
E\la4\times10^{19}\,$eV. At higher energies statistics is
insufficient to determine the variance of the depth of shower
maximum~\cite{Abraham:2010yv}. On the other
hand, HiRes observations are consistent with a light
composition above $\simeq1.6\times10^{18}\,$eV and up to
$\simeq5\times10^{19}\,$eV above which statistics is insufficient to
determine composition~\cite{Abbasi:2009nf}. This could indicate that statistics is
still too limited to draw firm conclusions or that the Northern and
Southern hemispheres are significantly different in terms of UHECR
composition. In addition, there are
significant uncertainties in hadronic cross sections, multiplicities and
inelasticities that can influence predicted air shower shapes
and none of the existing hadronic interaction models
consistently describes the shower depth and muon data of the Pierre Auger
experiment~\cite{Ulrich:2009hm,Ulrich:2010rg}. Note that the center of mass energy
for a UHECR interacting in the atmosphere reaches a PeV$=10^{15}\,$eV,
which is still a factor of a few hundred higher than the highest
energies reached in the laboratory, at the Large Hadron Collider (LHC)
at CERN. It is, therefore, not excluded that the true
chemical composition is light on both hemispheres and the UHECR data teaches
us something fundamental about hadronic interactions at energies unattainable in
the laboratory.

The question of chemical composition is linked to other observables
such as the UHECR spectrum. Unfortunately, the current statistics is
still insufficient to gain significant information on the chemical
composition from the observed spectrum. The flux suppression observed
above $\simeq4\times10^{19}\,$eV is qualitatively consistent
with either proton or nuclei heavier than carbon up to iron
nuclei~\cite{Allard:2008gj,Anchordoqui:2007tn,Anchordoqui:2007fi}. In
the latter case, the main energy loss process responsible for the
``cut-off'' is photo-disintegration on the CMB and infrared
backgrounds. It should be noted, however, that the observed flux
suppression could also be due to the intrinsic maximal acceleration
energies attained in the sources, although it would possibly be
somewhat of a coincidence that this energy should be close to the GZK
energy.

The UHECR chemical composition can in principle also be
tested independently with the flux of secondary cosmogenic
neutrinos~\cite{Anchordoqui:2007fi,Ahlers:2010fw,Kotera:2010yn}
and photons~\cite{Gelmini:2007jy,Hooper:2010ze}: These secondaries are
essentially produced by pion production on the constituent nucleons of a
nucleus with a given atomic number $A$. Therefore, if the maximal
acceleration energy $E_{\rm max}$ is not much larger than
$10^{21}\,$eV then for mass numbers $A$ approaching iron group nuclei,
the energy of the constituent nucleons will be below the GZK threshold
for pion production on the CMB and secondary gamma-ray and neutrino
production can only occur by interactions with the infrared background,
with a rate suppressed by the relative target photon number density
which is a factor of a few hundred. As a result, the cosmogenic
neutrino and photon fluxes depend strongly on injection spectrum,
maximal acceleration energy and chemical composition, but it may not
always be easy to break the resulting degeneracies.

Finally, the question of chemical composition of UHECRs is strongly linked with
the question of deflection angles in cosmic magnetic fields. In a
field with rms strength $B$ and coherence length $l_c$ the rms
deflection angle of a cosmic ray of energy $E$ and charge $Ze$
traveling a distance $d$ is given by~\cite{Waxman:1996zn}
\begin{eqnarray}\label{eq:deflec}
  \theta(E,d)&\simeq&\frac{(2dl_c/9)^{1/2}}{r_g}\\
  &\simeq&0.8^\circ\,
  Z\left(\frac{E}{10^{20}\,{\rm eV}}\right)^{-1}
  \left(\frac{d}{10\,{\rm Mpc}}\right)^{1/2}
  \left(\frac{l_c}{1\,{\rm Mpc}}\right)^{1/2}
  \left(\frac{B}{10^{-9}\,{\rm G}}\right)\,,\nonumber
\end{eqnarray}
where $r_L=E/(ZeB)$ is the Larmor radius. For an order of magnitude estimate for
the deflection angles in the Galactic magnetic field we use $l_c\sim100\,$pc,
$d\sim10\,$kpc, $B\sim3\,\mu$G gives
$\theta(E)\sim1^\circ\,Z(10^{20}\,{\rm eV}/E)$. Thus, protons around
the GZK cut-off, $E\sim60\,$EeV, will be deflected by a few degrees or
less, whereas iron nuclei can be deflected by several dozens of
degrees. This immediately raises the issue that the Galactic
magnetic fields are likely to destroy any possible correlation with the
local large scale structure in case of a heavy composition. Detailed
numerical simulations demonstrate that the relatively large
deflections of a heavy composition can considerably distort the images
of individual sources and even of the local large scale structure as a
whole~\cite{Giacinti:2010dk}.

Large scale extra-galactic magnetic fields (EGMF) are much less well
known than Galactic magnetic fields~\cite{EGMF-reviews}. One reason is that one of
the major detection methods for the EGMF, the Faraday rotation of the
polarization of radio emission from a distant source which is a
measure of the line of sight integral of the plasma density and the
parallel magnetic field component, is only sensitive to fields at a
given location stronger than $\sim0.1\mu$G. Fields below that strength
require much higher statistics data than currently available, but still have a strong
effect on UHECR deflection, as obvious from Eq.~(\ref{eq:deflec}). As
a statistical average over the sky, an all pervading EGMF is
constrained to be $\la3\times10^{-7}\,(l_c/{\rm
  Mpc})^{1/2}\,$G~\cite{Blasi:1999hu}.
Assuming an EGMF whose flux is frozen and follows the large scale
structure gives the more stringent limit $B\la10^{-9}-10^{-8}\,$G, but
the fields in the sheets and filaments can in this case be up to a
micro Gauss. This is also the scale which is routinely observed in
galaxy clusters which are the largest virialized structures in the
Universe. Beyond clusters at most hints exist on the EGMF, for example
in the Hercules and Perseus-Pisces superclusters~\cite{Xu:2005rb}.
It is expected, however, that in the future large scale radio
telescopes such as Lofar and SKA will improve observational
information on the EGMF in the large scale structure dramatically. We
note in this context that the EGMF in the voids is expected to be very
week and uncontaminated by astrophysical processes. This makes voids
excellent probes of relic seed magnetic fields from the early
Universe~\cite{Grasso:2000wj}. It is exciting that the non-observation at GeV
energies by Fermi of certain distant blazars that were seen at TeV
energies by HESS suggest a {\it lower limit} $E\ga3\times10^{-16}\,$G
on the EGMF in the voids~\cite{Neronov:2010gk}. This is because the TeV gamma-rays
seen by HESS would initiate electromagnetic cascades that should be
detectable by Fermi unless an EGMF of that strength deflects these
cascades into a diffuse halo around the source whose flux is then
below the Fermi sensitivity. However, void fields at that level are
not relevant for UHECR propagation.

\begin{figure}[ht!]
\includegraphics[scale=1.2]{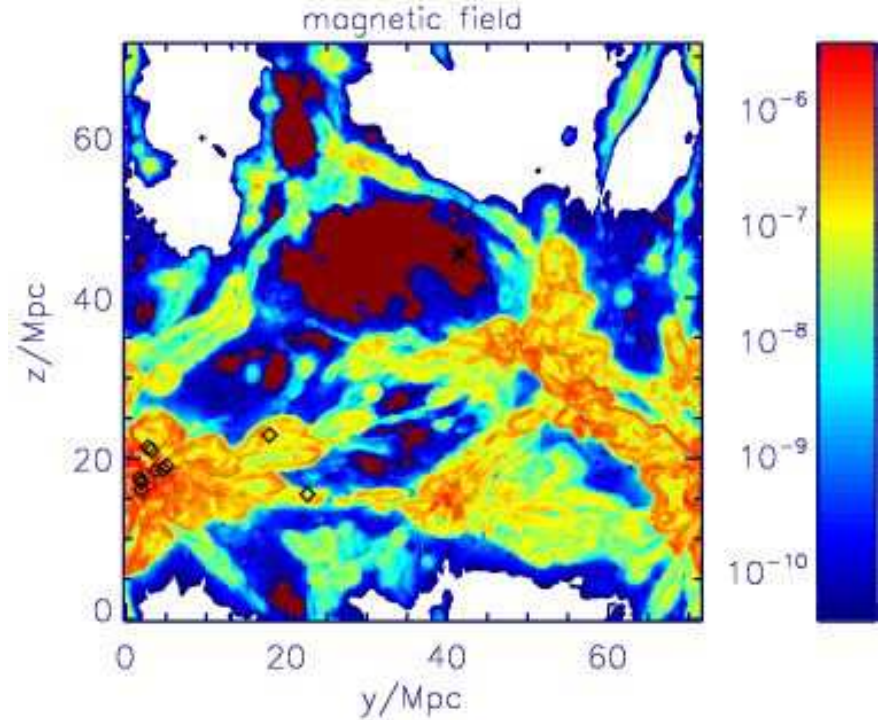}
%
%
\caption{A cross section through a typical large scale structure
  simulation such as the ones discussed in
Ref.~\cite{ryu,miniati}, on a scale of 70 Mpc in both directions. Ten sources
marked with diamonds in the environment of a massive galaxy cluster. The
black cross indicates the observer. The color contours represent the magnetic
field strength in units of Gauss, as indicated.}
\label{fig:1}       
\end{figure}

\begin{figure}[h!]
\sidecaption
\includegraphics[scale=.65]{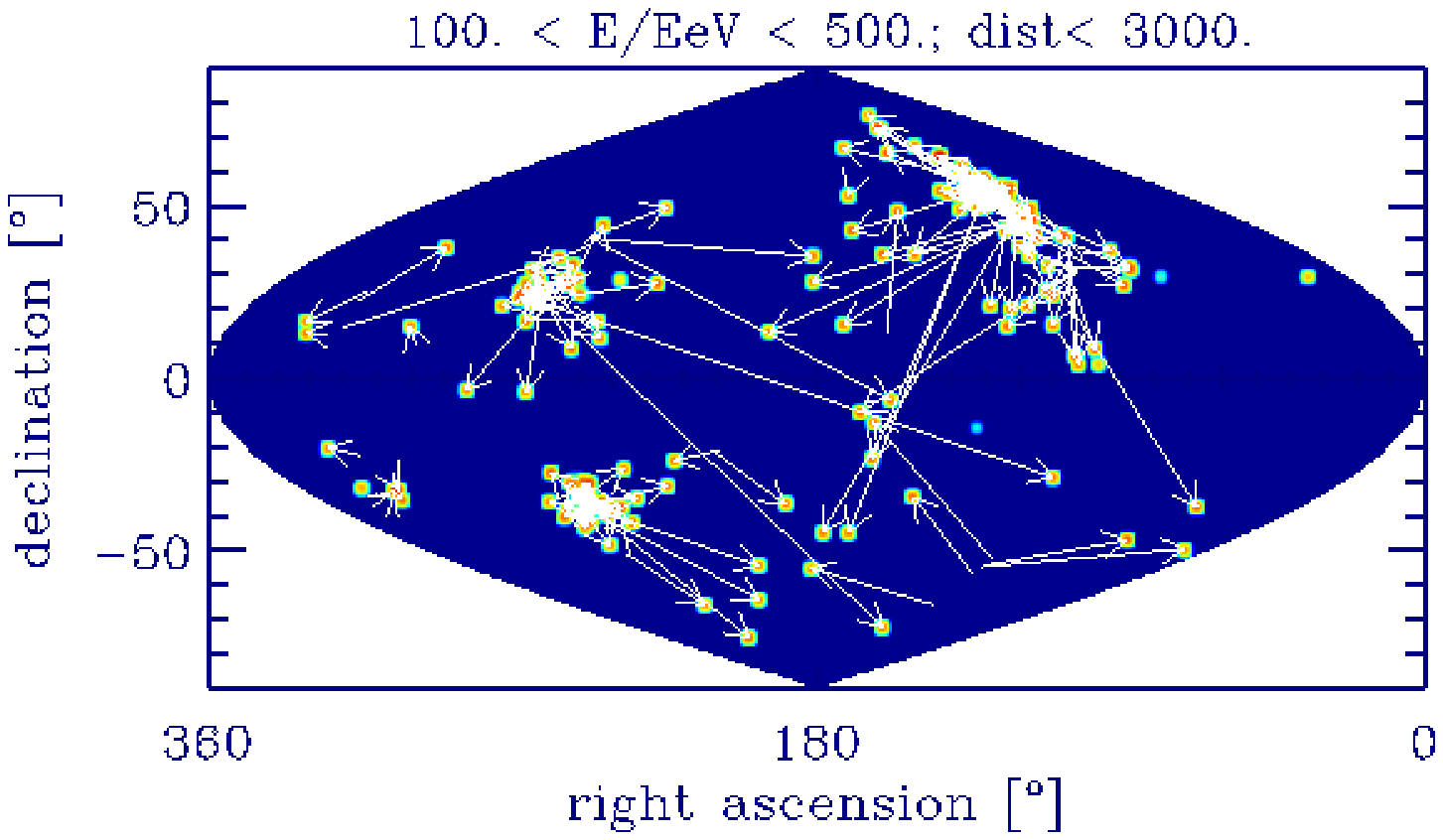}
\includegraphics[scale=.5]{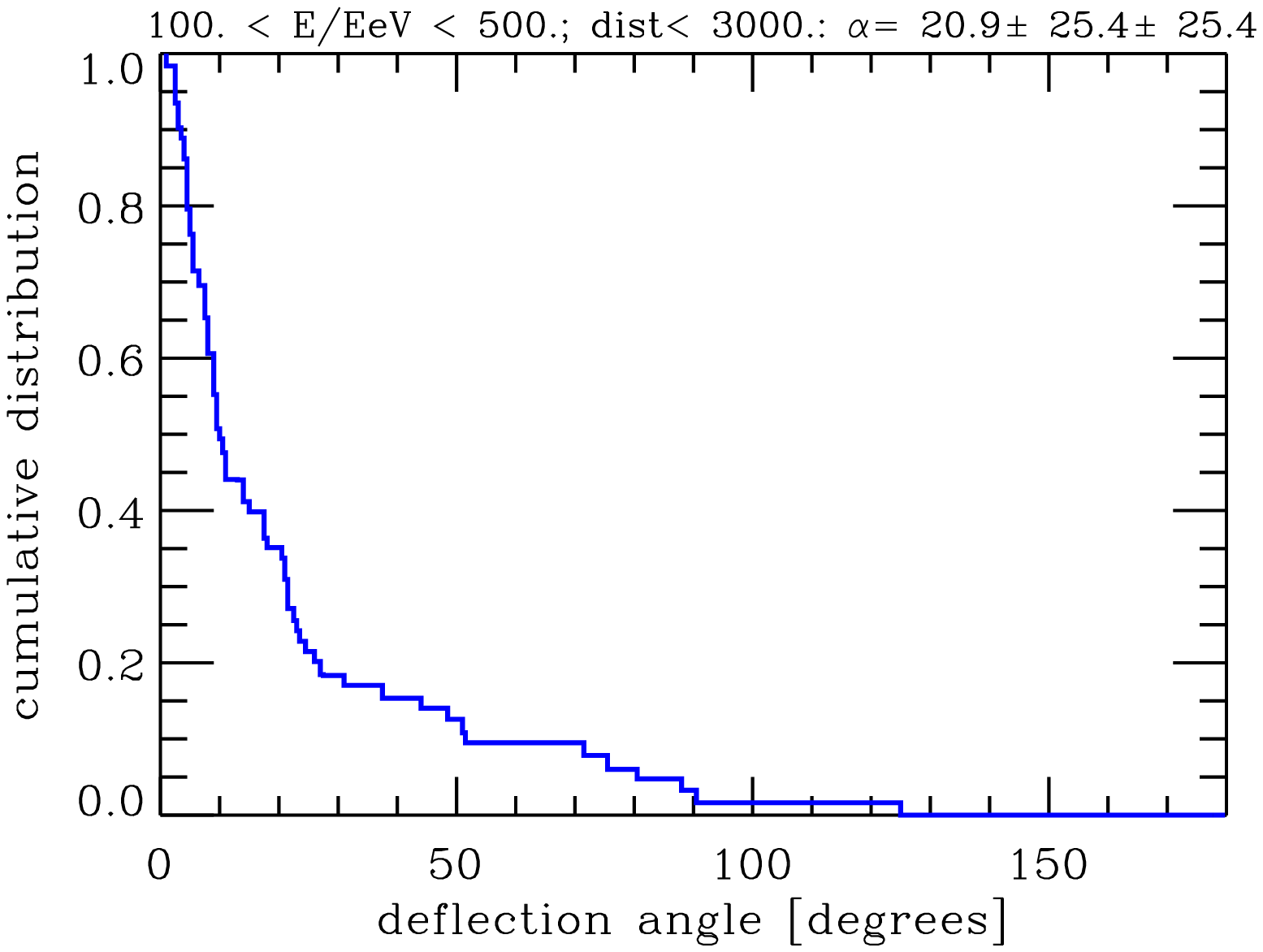}
%
%
\caption{Upper panel: Simulated arrival directions of UHECR above $10^{20}\,$eV in
a scenario where the sources shown in Fig.~\ref{fig:1} inject a pure
iron composition with an $E^{-2.2}$ spectrum and equal luminosity up to
$10^{22}\,$eV. The density of discrete sources in this simulation is
$\simeq2.4\times10^{-6}\,{\rm Mpc}^{-3}$ and the maximal distance the primaty
cosmic rays were allowed to propagate is $3000\,$Mpc. The arrows point
from the source to the detected event.
Lower panel: Distribution of deflection angles between arrival direction and
source position. The average deflection angle is $\simeq21^\circ$ with
a scatter of $\simeq26^\circ$.}
\label{fig:2}       
\end{figure}

As long as better observational information on the EGMF is not
available yet, one way of proceeding is to build models of the EGMF
using large scale structure simulations. Two major techniques for
doing this are a magnetohydrodynamic version of a constrained smooth particle
hydrodynamics code~\cite{EGMF-Dolag} and Eulerian grid-based hydro+n-body
codes~\cite{EGMF-Miniati}. The magnetic fields are followed
passively and are seeded either uniformly or around cosmic shocks
through the Biermann battery mechanism. The normalisation is then
constrained by the largest fields observed in galaxy
clusters. Alternatively, it has been assumed that the EGMF follows the
local vorticity and turbulent energy density of the matter~\cite{EGMF-Ryu}.
These numerical approaches agree on the fact that these fields
tend to follow the large scale galaxy structure, i.e. the fields tend to be strongest
around the largest matter concentrations. A cross section through one of
these simulations~\cite{ryu,miniati} is shown in Fig.~\ref{fig:1} (upper
panel). However, they disagree on certain aspects that are
relevant for UHECR deflection, most notably the filling factor distributions, i.e.
the fraction of space filled with EGMF above a certain strength, as a function of
that strength~\cite{Sigl:2004gi}. While this causes considerable differences in the 
size of the deflection angles
predicted between the source and the observed events, the deflections tend to be
{\it along and within} the cosmic large scale structure of the galaxy
distribution. This can be seen in Fig.~\ref{fig:2} where the upper
panel shows how the arrival directions relate to the source positions
on the sky and the lower panel shows the distribution of the deflection
angles between these two directions. In this scenario the deflected
UHECR arrival directions tend to follow arc-like structures that
result from deflections within the large scale cosmic filaments.
In other words, as long as the sources are not very nearby, the EGMF is unlikely
to deflect UHECRs out of the large scale structure since the fields in the voids
are very small. This means that the overall UHECR arrival direction
distribution arriving outside the Galaxy is likely to still correlate
with the local large scale structure even in the scenarios with large
EGMF, heavy nuclei and large deflection angles, although the events do in
general not point back to the sources. On the other hand, since deflections in the
Galactic field are unlikely to correlate with extragalactic
deflections, large deflections
of heavy nuclei in the Galactic field are expected to have a much stronger influence
on correlations with the local large scale structure.

\section{Testing fundamental symmetries: Lorentz-invariance and
cosmic gamma-rays}
\label{sec:3}
Both loop quantum gravity and string theory often break the Lorentz
symmetry or realize it in ways different from special
relativity. Typically, such effects manifest themselves through new
terms in the dispersion relation, the relation between energy $E$ and
momentum $p$ of a particle of mass $m$, that are suppressed by some
power $n$ of the Planck mass $M_{\rm Pl}$,
\begin{equation}\label{eq:dispersion}
  E^2=m^2+p^2\left[1+\eta\left(\frac{p}{M_{\rm Pl}}\right)^n\right]\,,
\end{equation}
where $\eta$ is a dimensionless number (we use natural units in which
the vacuum speed of light $c_0=1$). Such terms can modify both the
free propagation of particles and their interactions.

The propagation velocity now depends on energy in a different way than
in case of Lorentz invariance. In fact, in the relativistic limit
keeping only terms to first order in
$m^2$ and $\eta$, the group velocity for Eq.~(\ref{eq:dispersion}) is
\begin{equation}\label{eq:v}
  v=\frac{\partial E}{\partial p}\simeq1-\frac{m^2}{2E^2}+
  \frac{\eta}{2}(n+1)\left(\frac{E}{M_{\rm Pl}}\right)^n\equiv
  1-\frac{m^2}{2E^2}+\delta(E)\,,
\end{equation}
where $\delta(E)\equiv\eta(n+1)(E/M_{\rm Pl})^n/2$ is the deviation
from the Lorentz-invariant velocity. For photons, $m=0$, this can lead
to arrival time-delays between photons of different energies emitted
by GRBs or by flares of active galactic nuclei. Such time delays have indeed
been observed from space by Fermi LAT and Fermi GBM in the 10-100 GeV
region~\cite{Abdo:2009na} and from the ground, for example, by the
MAGIC telescope above 150 GeV~\cite{Albert:2007qk}. They have
been used to establish upper limits on the  Lorentz invariance
violating (LIV) terms. For $n=1$ these are typically of order one,
$|\eta|\la1$~\cite{Abdo:2009na}.

Furthermore, the kinematics of interactions can be modified which typically
happens when the LIV terms become comparable to the
particle rest mass, $E\ga E_{\rm cr}=(m^2M_{\rm Pl}^{n-2})^{1/n}$. As
a result, the larger the particle mass the higher the energy at
which LIV effects come into play. Therefore, TeV electrons and
positrons, but not protons, can be used to constrain $n=1$ LIV
effects~\cite{Maccione:2007yc}, and UHE protons are required to
obtain constraints on hadronic LIV terms with $n=2$ scaling. A
particularly interesting case is superluminal motion which occurs for
$\delta(E)>m^2/(2E^2)$ or $E>m/(2\delta)^{1/2}$, where for the general
case $\delta(E)$ is the difference of the LIV term for the particle
and the photon: At such energies a
charged particle would emit vacuum Cherenkov radiation, similar to the
motion of an ultra-relativistic charge in a medium with index of
refraction larger than one. The resulting rapid energy loss would
imply that particles can not reach such energies in astrophysical
environments. Their observation in turn allows to rule out the
corresponding LIV parameters.

The arguments above make it clear that LIV effects with $n\ge1$
increase with energy. The highest energies in Nature are observed in
high energy astrophysics, in particular TeV gamma-ray astrophysics
and UHE cosmic rays and neutrinos. There is thus a new field emerging
at the interface of quantum gravity phenomenology, string theory and
astrophysics. In fact, many of the LIV terms of the form of
Eq.~(\ref{eq:dispersion}) have already been strongly
constrained~\cite{LIV-reviews}. We mention in particular constraints
based on the flux suppression feature observed in UHECRs that is
consistent with the GZK effect: A tiny Lorentz invariance violation
with $\delta_\pi(E_\pi)-\delta_p(E_p)\ga5\times10^{-23}$ would lead to
a significant shift of the GZK feature and would thus be ruled
out~\cite{Stecker:2009hj}. In terms of $\eta$, for $n=2$, LIV effects
should thus be suppressed by a factor $\ga10^6$. LIV can also lead to
spontaneous decay, vacuum Cherenkov-radiation and modified
photo-disintegration reactions of very high energy nuclei, thereby
influencing UHECR chemical composition. This makes future UHECR
composition measurements also relevant for testing Lorentz invariance
violation~\cite{Saveliev:2011vw} .

In the following we will focus on photons for which the most important
interaction in an astrophysical and cosmological context is pair
production on low energy target photons ~\cite{Shao:2010wk}. The
highest energy photons we
know should be produced are the ones resulting from the decay of
$\pi^0$ mesons produced by the GZK effect. A certain fraction of the
UHECR flux should thus be photons. Due to pair production on the CMB
and infrared backgrounds and subsequent inverse Compton scattering of
the produced electrons and positrons an electromagnetic cascade
develops which quickly shifts the electromagnetic flux below the pair
production threshold on the CMB, $\simeq10^{15}\,$eV. As a result, the
expected photon fraction of the UHECR flux is rather small, less than
10\% around $10^{20}\,$eV and less than 1\% around
$10^{19}\,$eV~\cite{Gelmini:2007jy}. In fact, only experimental upper
limits are currently available consistent with the experimental
sensitivity~\cite{Collaboration:2009qb}.

However, a tiny Lorentz symmetry violation can inhibit pair production
such that the predicted UHE photon fraction would be much larger, of
the order of 20\% for $10^{19}\,{\rm eV}\la E\la10^{20}\,$eV, because
any photon produced by pion production, even at cosmological
distances, would only be subject to redshift and thus contribute to
the local UHE photon flux. This contradicts the observational upper
limits and can thus be used to constrain the LIV parameters in the
electromagnetic sector. The resulting constraints are very strong, in
fact much stronger than the ones obtained from arrival time dispersion
of gamma-rays from GRBs ~\cite{Abdo:2009na}: Typically, for LIV terms
suppressed to first order in the Planck scale, $n=1$, values
$|\eta|\ga10^{-14}$ are ruled out, whereas for second order
suppression, $n=2$, values $|\eta|\ga10^{-14}$ tend to be
constrained~\cite{Galaverni:2007tq,Galaverni:2008yj}. Since such
dimensionless coefficients would be expected to be of order one if they
are not forbidden by some symmetry, this suggests that LIV is most
likely absent altogether at first and second order suppression with
the Planck scale.

\section{Searching for new light states in electromagnetic
emission of astrophysical sources}
\label{sec:4}
Many extensions of the Standard Model of particle physics, in particular scenarios
based on supergravity or superstrings, predict a ``hidden sector'' of new particles
interacting only very weakly with Standard Model particles. Such scenarios do not
necessarily only contain Weakly Interacting Massive Particles (WIMPs), new heavy
states at the TeV scale and above some of which are candidates for the dark matter,
but often also predict Weakly Interacting Sub-eV Particles (WISPs)
that can couple to the photon field $A_\mu$~\cite{Jaeckel:2010ni}. The
most well-known examples include pseudo-scalar axions and axion-like
particles $a$ and hidden photons that mix kinetically with
photons.

Axion-Like Particles (ALPs) are described by a Lagrangian of the form
\begin{equation}\label{eq:axion}
  {\cal L}_{a\gamma}=\frac{1}{8\pi f_a}aF_{\mu\nu}\tilde F^{\mu\nu}+
  \frac{1}{2}m_a^2a^2=-\frac{1}{2\pi f_a}a{\bf E}\cdot{\bf B}+
  \frac{1}{2}m_a^2a^2\,,
\end{equation}
with $F_{\mu\nu}=\partial_\mu A_\nu-\partial_\nu A_\mu$ the electromagnetic
field tensor, $\tilde F^{\mu\nu}$ its dual, ${\bf E}$ and ${\bf B}$ the electric and
magnetic field strengths, respectively, $f_a$ a Peccei-Quinn like
energy scale and $m_a$ the axion mass. In addition, ALPs in general
have similar couplings to
gluons giving rise to mixing between axions and neutral pions
$\pi^0$. The actual axion was proposed to solve the strong CP-problem,
a problem of phase cancellation in quantum chromodynamics,
and exhibits a specific relation between coupling and mass,
$m_a\simeq0.6\,(10^{10}\,{\rm GeV}/f_a)\,$meV~\cite{Peccei:1977hh}.

A hidden photon field $X_\mu$ describes a hidden $U(1)$ symmetry group
and  mixes with the photon through a Lagrangian of the form
\begin{equation}\label{eq:hidden_photon}
{\cal L}_{X\gamma}=-\frac{1}{4}F_{\mu \nu}F^{\mu \nu}
 - \frac{1}{4}X_{\mu \nu}X^{\mu \nu}
+\frac{\sin\chi}{2} X_{\mu \nu} F^{\mu \nu}
+ \frac{\cos^2\chi}{2}m_{\gamma^\prime}^2 X_{\mu}X^\mu + j^\mu_{\rm em} A_{\mu}\,,
\end{equation}
where $X_{\mu\nu}$ is the hidden photon field strength tensor, $m_{\gamma^\prime}$ the
hidden photon mass and $\chi$ a dimensionless mixing parameter and
$j_{\rm em}^\mu$ is the electromagnetic current. Typical values for
the mixing parameter range from $\sim10^{-2}$ down to $10^{-16}$.

These couplings to photons can induce many interesting effects that are relevant
for astronomy and astrophysics: In the presence of electromagnetic fields, in
particular of magnetic fields, photons can
oscillate into axions and vice-versa, an effect known as 
Primakoff-effect~\cite{Pirmakoff:1951pj}. In fact, for a while this
possibility was even entertained as
a possible explanation of the disturbing observation that the
explosions of white dwarfs which can serve as ``standard candles''
because of their roughly constant explosion energy are dimmer than
expected in a decelerating Universe that would
otherwise lead to the conclusion that the expansion of the Universe
must accelerate~\cite{Riess:1998cb,Perlmutter:1998np}. Although
meanwhile this possibility is
basically excluded because it predicts other signatures, notably
distortions of the CMB, which have not been observed~\cite{Mirizzi:2005ng},
photon-ALPs mixing can still play a role at higher energies.

Photons can also oscillate into hidden
photons even in vacuum. These oscillations can be modified in the presence of
a plasma which gives the photons an effective mass whereas the WISP mass is
essentially unchanged. This can give rise to matter oscillations reminiscent
of the Mikheyev–Smirnov–Wolfenstein effect for neutrino 
oscillations~\cite{Wolfenstein:1977ue,Mikheev:1986gs}. In particular, even
if the mixing in vacuum is very small, one can have resonant conversions of
photons into WISPs within a plasma. Such photon conversions in vacuum and
in matter can have effects both within astrophysical sources and during propagation
of photons from the source to the observer.

The coupling of WISPs to photons and (in case of axions) also to fermions
can have an influence on the evolution and structure of astrophysical
objects. Due to their
weak coupling to ordinary matter, once produced, these hidden sector particles
can leave most objects without significant reabsorption, providing an efficient
cooling mechanism. This has lead, for example, to strong limits on axion masses
and couplings from the requirement that core-collapse supernovae should not
cool much faster than predicted if their cooling is dominated by neutrino
emission, in order to be consistent with the few neutrinos observed from the
cooling phase of SN1987A~\cite{Keil:1996ju}.

Even if the physics of the astronomical objects is not significantly
modified, the photon rates and spectra observable at Earth can be
influenced either within the source or during propagation to the
observer. A sensitive probe of photon-WISP
oscillations requires an as detailed an understanding of the emission
process as possible. In this context, one of the best understood
radiation sources in the Universe is the cosmic microwave background
(CMB). Its spectrum deviates from a perfect blackbody by less than
$\simeq10^{-4}$, distortions that have been measured by the COBE-FIRAS
experiment~\cite{Fixsen:1996nj}, and whose deviations from isotropy
are of the order of $10^{-5}$ and have themselves been measured at the
percent level by WMAP~\cite{Komatsu:2010fb}. This radiation
essentially comes from the surface of last scattering, at a distance
of a Hubble radius today, and any photon-WISP mixing at a level of
$\sim10^{-4}$ would induce a spectral distortion or an anisotropy in
conflict with the observations. This has lead to some of the strongest
limits on the parameters of Eqs.~(\ref{eq:axion})
and~(\ref{eq:hidden_photon}): For $10^{-9}\,{\rm eV}\la
m_a\la10^{-4}\,$eV one has $f_a\ga10^{11}(B_{\rm rms}/{\rm
  nG})\,10^{10}\,$GeV which strengthens to $f_a\ga10^{12}(B_{\rm rms}/{\rm
  nG})\,10^{11}\,$GeV for  $10^{-14}\,{\rm eV}\la
m_a\la10^{-11}\,$eV~\cite{Mirizzi:2009nq}. Since photon-ALP mixing
requires the presence of a magnetic field, the absence of significant
effects on the CMB imposes an upper limit on the combination $B_{\rm
  rms}/f_a$, with $B_{\rm rms}$ the rms large scale extra-galactic
magnetic field. Furthermore, requiring
the distortions of the CMB induced by photon-hidden
photon mixing to be smaller than the COBE-FIRAS limit leads
to a bound on the  mixing angle $\chi\la10^{-7}-10^{-5}$ for
hidden photon masses $10^{-14}\,{\rm eV}\la
m_{\gamma^\prime}\la10^{-7}\,$eV~\cite{Mirizzi:2009iz}. In contrast to
the case of ALPs, these contraints only depend on the vacuum mixing
angle $\chi$ since no external magnetic fields are necessary for
photon-hidden photon mixing.

Most other astrophysical sources are non-thermal in nature and thus
much less well understood. This is the case in particular for X-ray
and gamma-ray sources. Still, if the photon spectra from these objects
can be well approximated by power laws, photon-ALPs mixing can induce
steps in the spectra that may be detectable. Depending on the strength
of magnetic field within the sources, for ALP masses
$m_a\sim10^{-6}\,$eV significant effects on spectra
between keV and TeV energies can occur for
$f_a\la10^{13}\,$eV~\cite{Hooper:2007bq,Hochmuth:2007hk}. These
effects are complementary and potentially more sensitive compared to
more direct experimental bounds the best of which come from {\it
  helioscopes}: Photons from the
sun are converted to ALPs in the solar magnetic field which in turn can be
reconverted to photons in an artificial magnet in front of a telescope
on Earth which then detects these photons. For $m_a\la0.02\,$eV the
CERN Axion Solar Telescope (CAST) experiment provided the strongest
constraint, $f_a\ga10^{10}\,$GeV~\cite{Arik:2008mq}.

Since photon-ALP mixing is energy dependent, ALP signatures are best
revealed when comparing luminosities at different energies. In
particular, it has been pointed out that the {\it scatter} of
correlations of luminosities in different energy bands deviates from a
Gaussian if photon-ALP mixing occurs. In fact, considerable deviations
from Gaussian scatters
have recently been found in the correlations between the luminosities
of AGNs in the optical/UV and X-rays~\cite{Burrage:2009mj}. If these
sources are located in galaxy clusters which are known to contain
magnetic fields of micro Gauss strength, photon-ALP mixing could
explain this observation if $m_a\ll10^{-12}\,$eV and
$f_a\la10^{10}\,$GeV. In this case, almost energy
independent photon-ALP mixing would occur at energies above
$\simeq2\,$keV, whereas the mixing would be highly energy dependent at
energies $\ll0.5\,$keV, thereby inducing non-Gaussian
correlations. Similar effects would occur with photon-ALP conversion
in magnetic fields within AGNs if $m_a\ll10^{-7}\,$eV and
$f_a\simeq3\times10^8\,$GeV. It has been pointed out, however, that
the scatter in the correlation between optical and X-ray
luminosities observed in AGNs can also be explained by X-ray
absorption~\cite{Pettinari:2010ay}.

Another possible signature for photon mixing with a new light state
has been discussed in the context of high energy gamma-ray
observations by the ground-based telescopes MAGIC, H.E.S.S., VERITAS
and CANGAROO-III. The absorption of such gamma-rays
in the infrared background appears weaker than expected based on
models for the infrared
background~\cite{Aharonian:2005gh,Stecker:2007jq}, although
this is currently
inconclusive~\cite{Aliu:2008ay,Costamante:2009gz}. If gamma-ray absorption is
indeed weaker than computed for the real infrared background, this
could be explained if part of
the gamma-rays are converted into ALPs around the source which in turn
are reconverted into gamma-rays in the Galactic magnetic
field~\cite{Simet:2007sa}. This works for ALP parameters $10^{-10}\,{\rm eV}\la
m_a\la10^{-8}\,$eV and $f_a\sim10^9\,$GeV
Alternatively, conversion and re-conversion could be induced by the
EGMF if $m_a\la10^{-10}\,$eV and $5\times10^{10}\,{\rm
  GeV}\la f_a\la10^{18}\,$GeV~\cite{DeAngelis:2007dy,DeAngelis:2008sk}.
A recent detailed study on these effects has been performed in
Ref.~\cite{SanchezConde:2009wu}. We note, however, that an apparently
reduced absorption of $\gamma-$rays from high redshift sources can
also be explained if these $\gamma-$rays are produced near Earth by
primary TeV-PeV {\it cosmic rays} from the same source which interact much
less frequently with the low energy target photons than than TeV
$\gamma-$rays~\cite{Essey:2010er}. This is possible provided that
cosmic ray deflection is sufficiently small, corresponding to large
scale EGMFs of strength $B\la3\times10^{-14}\,$G~\cite{Essey:2010nd}.

\section{Conclusions}
\label{sec:5}
In this contribution we have discussed three examples in which
astronomy plays an interdisciplinary role at the intersection with the
neighboring scientific fields of cosmology and particle physics: The
nature and origin of the highest energy particles observed in Nature,
tests of the Lorentz symmetry which is one of the pillars of
modern science tiny breakings of which may yield fundamental insights
into Nature and may lead to observable effects at the highest energies,
and, at the opposite end of the energy scale, the mixing of photons
with new light states such as axion-like particles or hidden
photons. While this list is certainly not exhausting and does not
include other important topics such as the search for dark matter, it
hopefully gives an idea about the role of interdisciplinarity in
astronomy. With the first results coming in from the Large Hadron
Collider, the most powerful existing particle physics experiment in
terms of energy and luminosity, new levels of
cross-fertilization between astronomy and particle physics are
expected for the near future.

\begin{acknowledgement}
This work was
supported by the Deutsche Forschungsgemeinschaft through the
collaborative research centre SFB 676 Particles, Strings and the Early
Universe: The Structure of Matter and Space-Time and by the State of
Hamburg, through the Collaborative Research program Connecting
Particles with the Cosmos within the framework of the
Landesexzellenzinitiative (LEXI).
\end{acknowledgement}
%

%
%
%

\end{document}